\documentstyle[prl,aps,twocolumn]{revtex}
\begin{document}
\draft
\title{Thickness-Magnetic Field{\bf \ }Phase Diagram at the
Superconductor-Insulator Transition in 2D}
\author{N. Markovi\'{c}, C. Christiansen and A. M. Goldman}
\address{School of Physics and Astronomy, University of Minnesota, Minneapolis,\\
MN 55455, USA}
\date{May 1,1998}
\maketitle

\begin{abstract}
The superconductor-insulator transition in ultrathin films of amorphous Bi
was tuned by changing the film thickness, with and without an applied
magnetic field. The first experimentally obtained phase diagram is mapped as
a function of thickness and magnetic field in the T=0 limit. A finite size
scaling analysis has been carried out to determine the critical exponent
product $\nu z,$ which was found to be $1.2\pm 0.2$ for the zero field
transition, and $1.4\pm 0.2$ for the finite field transition. Both results
are different from the exponents found for the magnetic field tuned
transition in the same system, $0.7\pm 0.2.$
\end{abstract}


Superconductor-insulator (SI) transition in ultrathin films of metals is
believed to be a continuous quantum phase transition \cite{Sondhi} which can
be traversed by changing a parameter such as disorder, film thickness,
carrier concentration or the applied magnetic field \cite{Fisher d,Fisher B}%
. The scaling theory and a phase diagram for a two-dimensional system as a
function of disorder and magnetic field was postulated by Fisher {\it et al}%
. \cite{Fisher B,Fisher g}, based on the assumption that this transition can
be fully described in terms of a model of interacting bosons, moving in the
presence of disorder. The dirty boson problem has been extensively studied
by quantum Monte Carlo simulations \cite{Scalettar,Wallin,Cha,Kisker},
real-space renormalization group calculations \cite{Singh,Zhang}, strong
coupling expansion \cite{Freericks} and in other ways \cite
{Ma,Fradkin,Herbut}, but there is still some disagreement as to the
universality class of the transition. Conflicting experimental evidence
suggests that the bosonic model might be relevant \cite{Hebard}, but does
not give the full picture \cite{Dynes}. An alternative model of interacting
electrons has also been proposed \cite{Trivedi}. Experimentally, the
thickness tuned transition has been studied in the context of the scaling
theory in zero magnetic field \cite{Liu}. In the present work, for the first
time, the SI transition was tuned by systematically changing the film
thickness in a finite magnetic field. This allows us to map a phase diagram
as a function of thickness and magnetic field in the T=0 limit and to
determine the critical exponents using a finite size scaling analysis at
different fields. The results suggest that this transition is similar to the
zero field transition, but the exponent is different from that of the
magnetic-field tuned transition studied on the same set of films \cite
{Markovic}.

The ultrathin Bi films were evaporated on top of a 10 \AA\ thick layer of
amorphous Ge, which was pre-deposited onto a 0.75 mm thick single-crystal of 
$SrTiO_{3}$ $(100)$. The substrate temperature was kept below 20 K during
all depositions and all the films were grown $in$ $situ$ under UHV
conditions ($\sim 10^{-10}$ Torr). Under such circumstances, successive
depositions can be carried out without contamination to increase the film
thickness gradually in increments of $\sim 0.2\AA $. Film thicknesses were
determined using a previously calibrated quartz crystal monitor. Films
prepared in this manner are believed to be homogeneous, since it has been
found that they become connected at an average thickness on the order of one
monolayer \cite{Strongin}. Resistance measurements were carried out between
the depositions using a standard dc four-probe technique with currents up to
50 nA. Magnetic fields up to 12 kG perpendicular to the plane of the sample
were applied using a superconducting split-coil magnet.

The evolution of the temperature dependence of the resistance as the film
thickness changes is shown on Fig. \ref{R vs T for d}. The thinnest films
show an exponential temperature dependence of the resistance at low
temperatures, consistent with variable range hopping, which crosses over to
a logarithmic behavior for thicker films \cite{Mack}. At some critical
thickness, $d_{c},$ the resistance is independent of temperature, while for
even thicker films it decreases rapidly with decreasing temperature,
indicating the onset of superconductivity. The critical thickness can be
determined by plotting the resistance as a function of thickness for
different temperatures (inset of Fig. \ref{R vs T for d}) and identifying
the crossing point for which the resistance is temperature independent, or
by plotting $dR/dT$ as a function of thickness at the lowest temperatures
and finding the thickness for which $(dR/dT)=0$.

In the zero temperature quantum critical regime the resistance of a two
dimensional system is expected to obey the following scaling law \cite
{Sondhi,Fisher g}:

\begin{equation}
R(\delta ,T)=R_{c}\ f(\delta T^{-1/\nu z})  \label{one}
\end{equation}
Here $\delta =d-d_{c}$ is the deviation from the critical thickness, $R_{c}$
is the critical resistance at $d=d_{c}$, $f(x)$ is a universal scaling
function such that $f(0)=1$, $\nu $ is the coherence length exponent, and $z$
is the dynamical critical exponent. We rewrite Eq.\ \ref{one} as $R(\delta
,t)=R_{c}f(\delta t)$, where $t\equiv T^{-1/\nu z},$ and treat the parameter 
$t(T)$ as an unknown variable which is determined at each temperature to
obtain the best collapse of all the data. The exponent $\nu z$ is then found
from the temperature dependence of $t$, which must be a power law in
temperature for the procedure to make sense. This scaling procedure does not
require detailed knowledge of the functional form of the temperature or
thickness dependence of the resistance, or prior knowledge of the critical
exponents. It is simply based on the data which includes an independent
determination of $d_{c}$.

The collapse of the resistance data as a function of $\delta t$ in zero
field is shown in Fig. \ref{d collapse}. The critical exponent product $\nu
z $, determined from the temperature dependence of the parameter t (inset of
Fig. \ref{d collapse}), is found to be $\nu z=1.2\pm 0.2.$ This result is in
agreement with the predictions of Ref. \cite{Fisher d,Fisher B}, from which
z=1 would be expected for a bosonic system with long range Coulomb
interactions independent of the dimensionality, and $\nu \geq 1$ in two
dimensions for any transition which can be tuned by changing the strength of
the disorder \cite{Chayes}. A similar scaling behavior has been found in
ultrathin films of Bi by Liu et al. \cite{Liu}, with the critical exponent
product $\nu z\approx 2.8$ on the insulating side and $\nu z\approx 1.4$ on
the superconducting side of the transition. The fact that $\nu z$ was found
to be different on the two sides of the transition raises the question of
whether the experiment really probed the quantum critical regime. We believe
that the scaling was carried out too deep into the insulating side, forcing
the scaling form (Eq. \ref{one}) on films which were in a fundamentally
different insulating regime \cite{Mack}. Such films should not be expected
to scale together with the superconducting films, hence the discrepancy on
the insulating side of the transition. In the present work, the measurements
were carried out at lower temperatures than previously studied and with more
detail in the range of thicknesses close to the transition. We were able to
scale both sides of the transition with $\nu z\approx 1.2,$ which is close
to the value obtained by Liu et al. on the superconducting side of the
transition. Our result is also in very good agreement with the
renormalization group calculations \cite{Singh,Zhang,Herbut}, and close to
that found in Monte Carlo simulations by Cha and Girvin \cite{Cha}, S\o
rensen et al. \cite{Wallin} and Makivi\'{c} et al. \cite{Scalettar}.

In addition to the above, the magnetoresistance as a function of temperature
and magnetic field was measured for each film. By sorting the
magnetoresistance data, one can probe the thickness-tuned
superconductor-insulator transition in a finite magnetic field, which has
not been studied before. The same analysis as described above was carried
out for several fixed magnetic fields, ranging from 0.5 kG up to 7 kG. The
normalized resistance data as a function of the scaling variable for six
different values of the magnetic field shown on Fig.\ref{d SIT in B} all
collapse on{\it \ a single curve}, which suggests that the scaling function
is indeed universal. The critical exponent product determined from the
temperature dependence of the parameter t (inset of Fig. \ref{d SIT in B})
is found to be $\nu z=1.4\pm 0.2,$ apparently independent of the magnetic
field. An applied magnetic field is generally expected to change the
universality class of the transition, since it breaks the time reversal
symmetry. We find, however, that the critical exponent product $\nu z$ for
the thickness driven SI transition in a finite magnetic field is very close
to that obtained for a zero-field transition, given the experimental
uncertainties. This result is in agreement with Monte Carlo simulations of
the (2+1)-dimensional classical XY model with disorder by Cha and Girvin 
\cite{Cha}, which find $\nu z\approx 1.07$ for the zero-field transition and 
$\nu z\approx 1.14$ in a finite magnetic field. The critical resistance $%
R_{c}$ at the transition is non-universal, as it decreases with increasing
magnetic field.

Furthermore, once a magnetic field is applied and the time-reversal symmetry
broken, the thickness-tuned transition in the field is expected to be in the
same universality class as the transition which is tuned by changing the
magnetic field at a fixed film thickness. The magnetic-field tuned
transition was studied on the same set of films used in this study \cite
{Markovic}, which allows a direct comparison of the critical exponents,
without having to worry about differences in the microstructure between
different samples. The resistance as a function of temperature for five
selected films from Fig. \ref{R vs T for d} was studied in magnetic fields
up to 12kG applied perpendicular to the plane of the sample. The critical
exponent product was determined using the method described above, but with
the magnetic field as the tuning parameter rather than the film thickness.
It was found to be $\nu _{B}z_{B}=0.7\pm 0.2$, independent of the film
thickness, which strongly suggests a universality class different from that
of the thickness-tuned transition, both in zero-field and in the field.\
This result does not agree with the predictions based on the model of
interacting bosons in the presence of disorder \cite{Fisher B,Fisher g}. It
does, however, agree with what might be expected from a similar model
without disorder \cite{Kisker,Cha}. The details of the analysis and this
unexpected value of the critical exponent product, as well as the discussion
of its disagreement with previous determinations \cite{Yazdani}, are
reported elsewhere \cite{Markovic}.

The fact that the SI transition was traversed by changing both the thickness
and the magnetic field independently on the same set of films allows us to
determine the phase diagram as a function of thickness and the magnetic
field in T=0 limit, which is shown in Fig. \ref{phase diagram}. The films
characterized by parameters which lie above the phase boundary are
''superconducting'' ($\delta R/\delta T<0$ at finite temperatures), and the
ones bellow it are ''insulating'' ($\delta R/\delta T>0$ at finite
temperatures). The phase boundary itself follows a power law: $
d_{c}(B)-d_{c}(0)\propto B_{c}^{x},$ where $x\approx 1.4$. Our results imply
that the critical exponent product $\nu z$ depends on whether this phase
boundary is crossed vertically (changing the thickness at a constant
magnetic field), in which case $\nu z\approx 1.4,$ or horizontally (changing
the magnetic field at a fixed thickness), in which case $\nu _{B}z_{B}%
\approx\ 0.7$. In other words, relatively weak magnetic fields which are
experimentally accessible to us do not significantly change the universality
class of the thickness-tuned transition, but the magnetic-field tuned
transition is in a different universality class from the thickness-tuned
transition in $B\neq 0$. In the absence of a detailed theory, we can only
speculate on the origins of this surprising result.

The first important issue we wish to discuss is the role of the film
thickness as the control parameter. Adding metal sequentially to a
quench-condensed film has been shown to decrease the disorder, since the
increased screening smooths the random potential seen by the electrons. It
presumably also increases the carrier concentration, which in the presence
of an attractive electron-electron interaction might result in an increased
Cooper pair density. 

Increasing the film thickness might therefore be thought of as adding Cooper
pairs, which condense at some critical density. In a similar way, an applied
magnetic field adds vortices, which behave as point particles and also
condense at some critical density, making the system insulating. However,
this symmetry between charges and vortices is not perfect, since Cooper
pairs interact as 1/r and vortices interact logarithmically \cite{Keldysh}. 

If the mechanism responsible for the localization in the magnetic-field
tuned transition is different from that of the thickness-tuned transition,
than having a non-zero magnetic field may not play a major role in the
thickness-tuned transition. Also, the correlation length associated with the
thickness-tuned transition would then be different from that associated with
the magnetic-field tuned transition.{\bf \ }The disorder might be important
in one case and not in the other, depending on how these correlation lengths
compare to the length scale which characterizes the disorder.

A second difference between the thickness-tuned and the field-tuned
transitions may be the nature of disorder itself. In the field-tuned
transition the geometry of the film is fixed, and the disorder does not
change. In the disorder-tuned transition, each film in the sequence of films
will have a slightly different microstructure, so that the disorder may have
to be averaged over the different configurations. It has been suggested
recently \cite{Pazmandi} that the nature of the disorder averaging might
play an important role in determining the critical exponents.

Another possibility is that the localization transition is enhanced by
percolation effects \cite{Ramakrishnan} as the film thickness is tuned. This
approach takes into account local fluctuations of the amplitude of the
superconducting order parameter, which are routinely neglected in the
scaling theory and the numerical simulations. Percolation of islands with
strong amplitude fluctuations might change the localization exponents
obtained from the scaling theory \cite{Ramakrishnan}. The role of
percolation has also been emphasized in recent discussions of low
temperature transport in these systems\cite{Shimshoni}.

Finally, one cannot exclude the possibility that to access the quantum
critical region these measurements need to be carried out at much lower
temperatures or at high frequencies \cite{Sachdev}.

We gratefully acknowledge useful discussions with A. P. Young, S. Sachdev
and S. L. Sondhi. This work was supported in part by the National Science
Foundation under Grant No. NSF/DMR-9623477.

\begin{figure}[tbp]
\caption{Resistance per square as a function of temperature for a series of
bismuth films with thicknesses ranging from 9\AA\ (top) to 15\AA\ (bottom).
Inset: Resistance as a function of thickness for the same set of films close
to the transition at low temperature. Different curves represent different
temperature, ranging from 0.14K to 0.40K.}
\label{R vs T for d}
\end{figure}

\begin{figure}[tbp]
\caption{Resistance per square as a function of the scaling variable, t%
%TCIMACRO{\TEXTsymbol{\vert} }
%BeginExpansion
\mbox{$\vert$}%
%EndExpansion
d-d$_{c}|$, for seventeen different temperature, ranging from 0.14K to 0.5K.
Here $t=T^{-1/\nu z}$ is treated as an adjustable parameter to obtain the
best collapse of the data. Different symbols represent different
temperature. Inset: Fitting the temperature dependence of the parameter t to
a power law determines the value of $\nu z.$}
\label{d collapse}
\end{figure}

\begin{figure}[tbp]
\caption{Normalized resistance per square as a function of the scaling
variable, $t|d-d_{c}|$, at different temperature, ranging from 0.14K to
0.5K. Different symbols represent different magnetic fields, ranging from
0.5kG-7kG. Here $t=T^{-1/\nu z}$ is treated as an adjustable parameter to
obtain the best collapse of the data, and R$_{c}$ is the resistance at $%
d=d_{c}$. Inset: Fitting the temperature dependence of the parameter t to a
power law determines the value of $\nu z.$}
\label{d SIT in B}
\end{figure}

\begin{figure}[tbp]
\caption{The phase diagram in the d-B plane in the T=0 limit. The points on
the phase boundary were obtained from disorder driven transitions
(triangles) and magnetic field driven transitions (circles). The solid line
is a power law fit. The values of the critical exponent product are shown
next to the arrows giving the direction in which the boundary was crossed.
Here d$_{c}$ is taken to be the critical thickness in zero field.}
\label{phase diagram}
\end{figure}

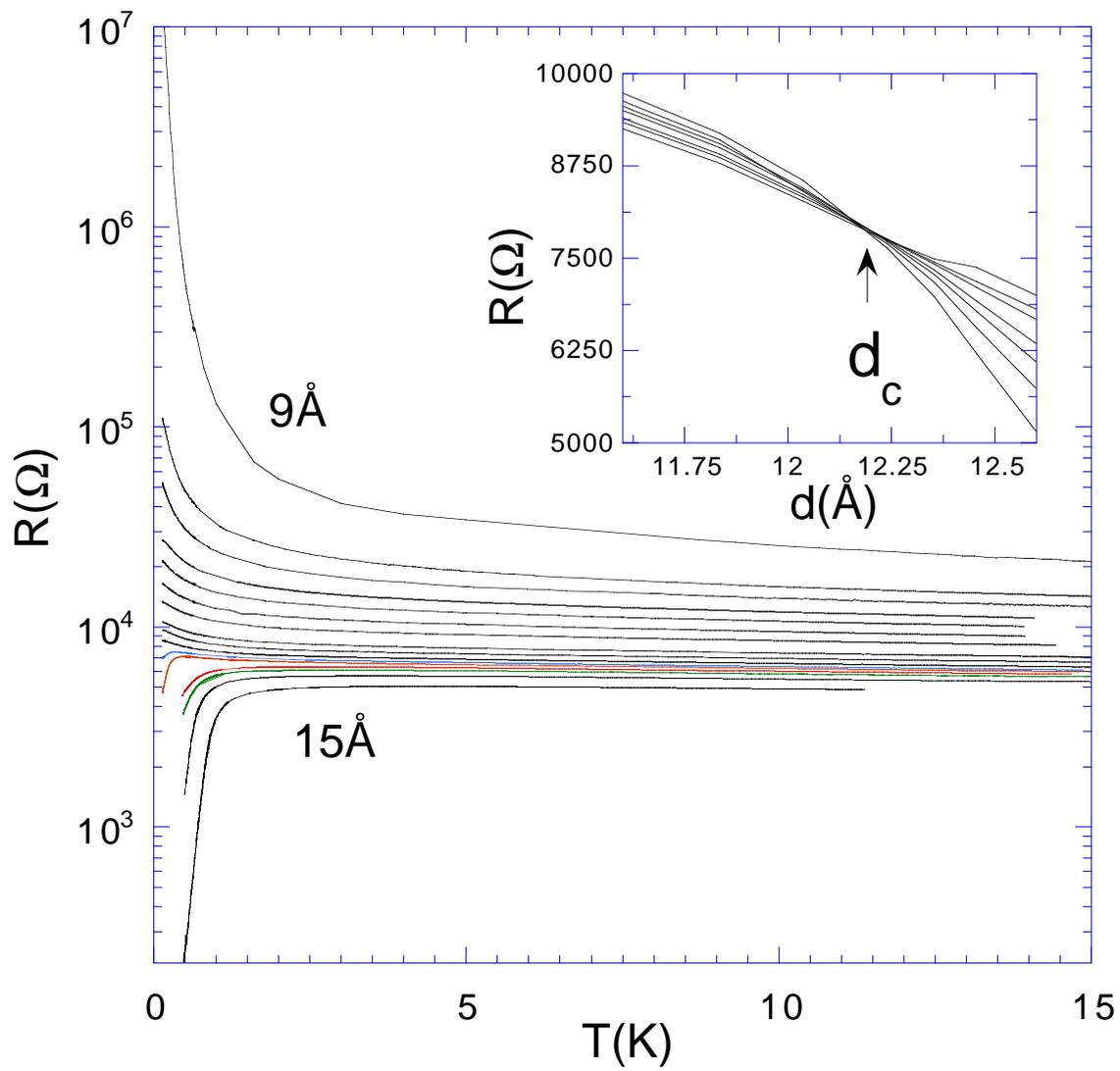
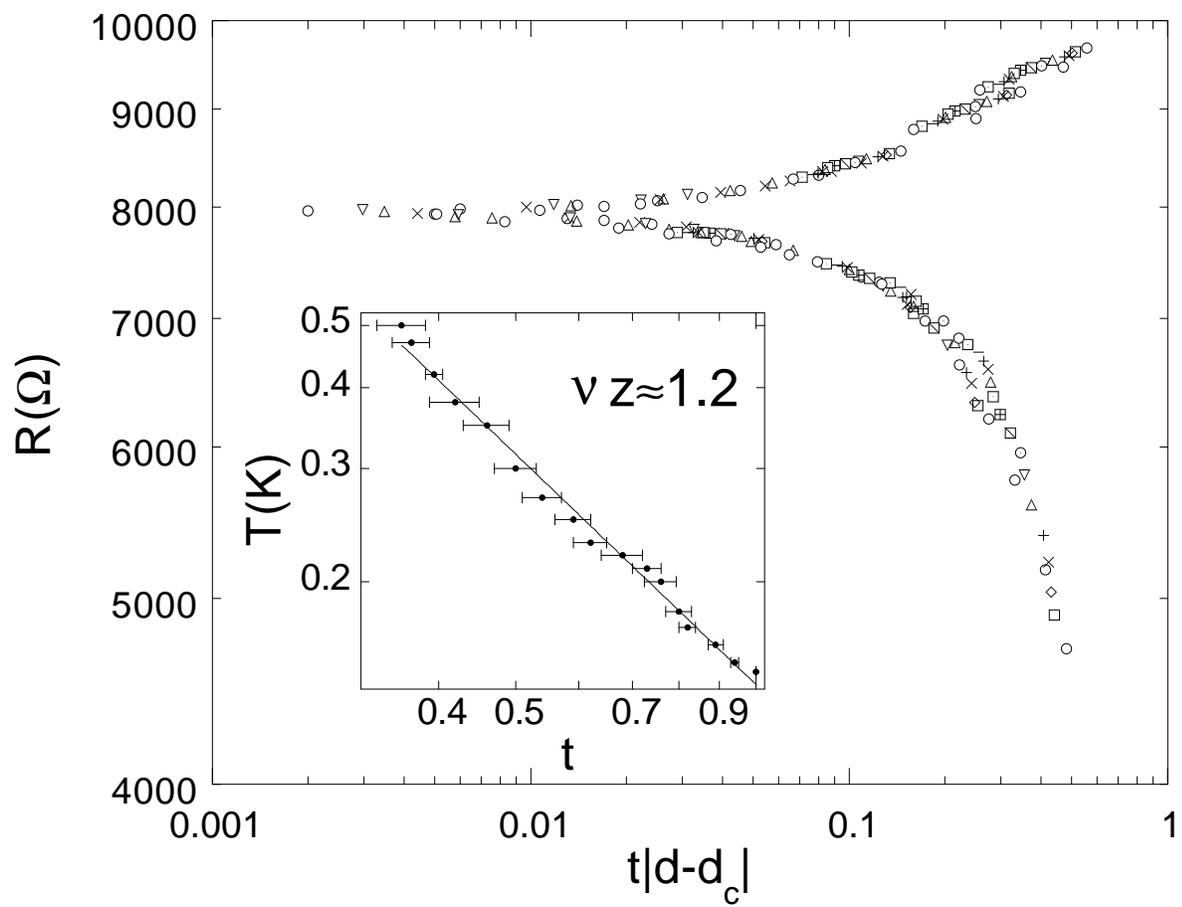
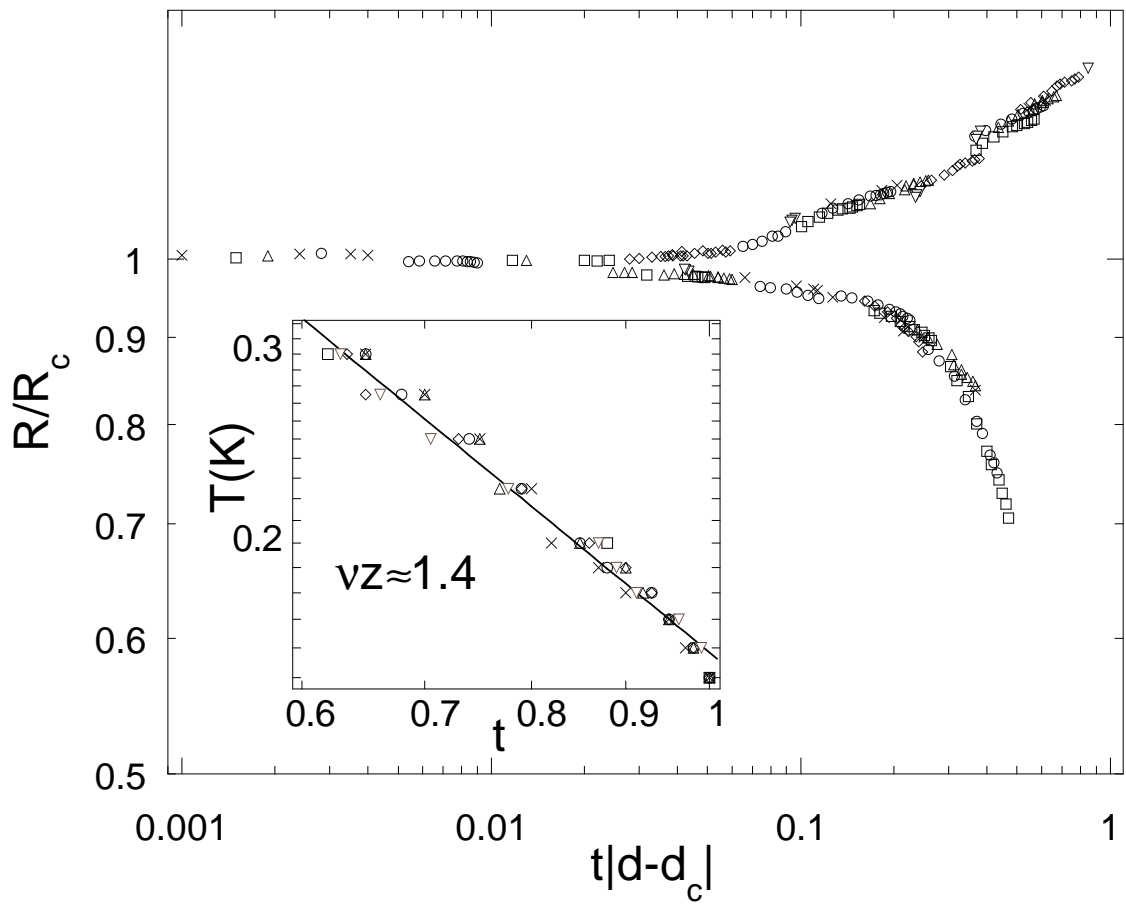
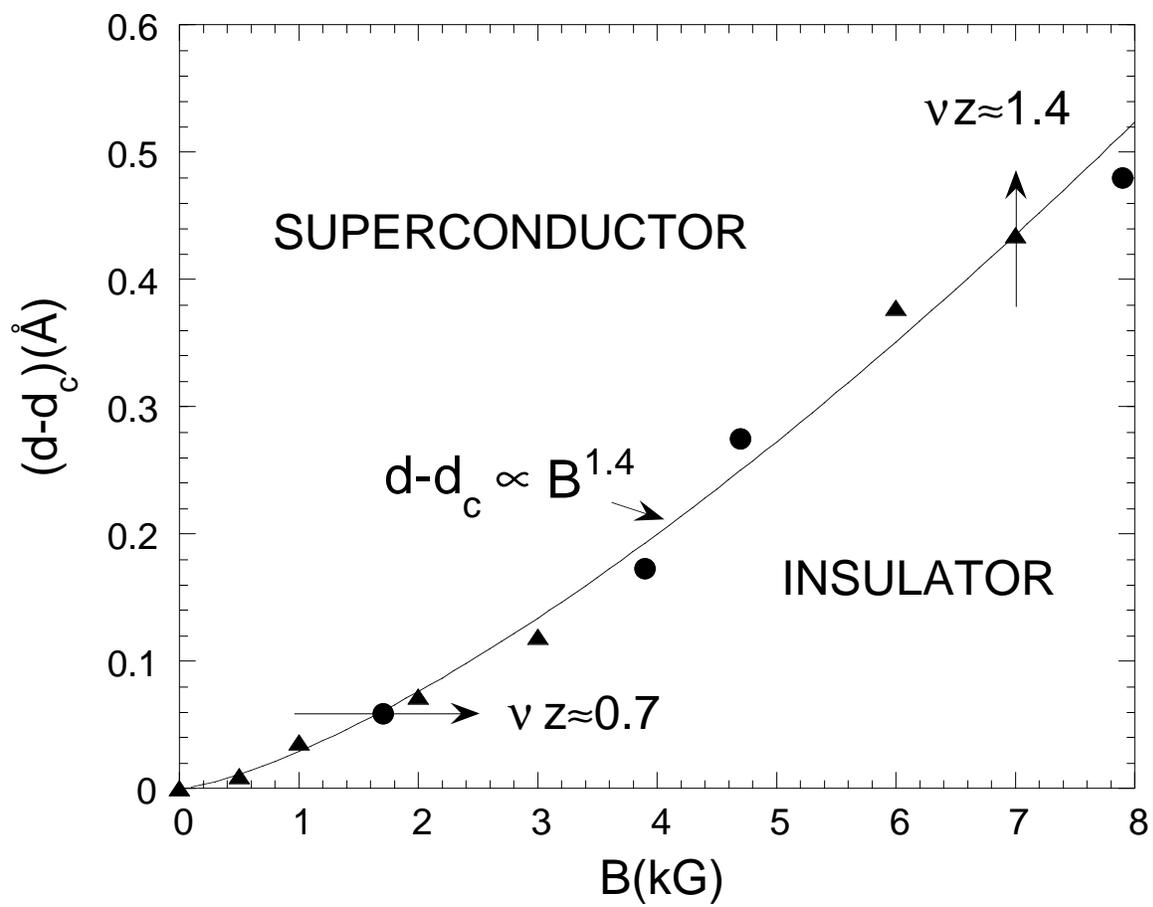
\end{document}